\documentclass[aps,prl,twocolumn,showpacs,amsmath,amssymb,superscriptaddress]
{revtex4}
\usepackage{graphicx}
\usepackage{bm}
\usepackage{psfrag}
\def\avg#1{\langle#1\rangle}

\def\be{\begin{equation}}       \def\ee{\end{equation}}
\def\bea{\begin{eqnarray}}      \def\eea{\end{eqnarray}}

\def\PRA{Phys. Rev. A~}
\def\PRB{Phys. Rev. B~}

\def\RMP{Rev. Mod. Phys. ~}
\def\PRL{Phys. Rev. Lett.~}
\begin{document}

\title{Josephson Diode}
\author{Jiangping Hu}
\affiliation{Department of Physics, Purdue University, West
Lafayette, IN  }
\author{ Congjun Wu}
\affiliation{Kavli Institute for Theoretical Physics, University of
California , Santa Barbara, CA}
\affiliation{Department of Physics, University of California, San Diego,
CA}
\author{Xi Dai}
\affiliation{Institute of Physics, Chinese Academy of Sciences,
Beijing, China}
 \affiliation{Department of Physics, and Center of
Theoretical and Computational Physics, the University of Hong Kong,
Hong Kong}

\begin{abstract}
We propose a new type of Josephson junction formed by two
superconductors close to  the superconductor-Mott-insulator
transition, one of which is doped with holes and the other is doped
with electrons.
A self-organized Mott-insulating depletion region is formed
at the interface between two superconductors, giving rise to an asymmetric
response of current to the external voltage.
The collective excitations of the depletion region
result in a novel phase dynamics that can be
measured experimentally in the noise spectrum of the
Josephson current.
\end{abstract}

\pacs{85.25.-j, 74.81.Fa 73.43.Nq} \maketitle
The Josephson effect is one of the most fundamental effects associated with
the superconducting phase, regardless of differences among various
superconducting materials.  In strongly correlated electron systems,
{
many other quantum phases exist in the vicinity of the
superconducting phase, which gives rise to many salient features in
Josephson junctions.
For example, a long range proximity effect in
superconducting/antiferromagnetic/superconducting (SAS) junctions
was predicted \cite{demler1} based on the competition between the
superconducting and antiferromagnetic phases in high temperature
superconductors \cite{zhang,project,subir,Eugene}.
An abrupt change of the Josephson critical current was predicted
in the junction arrays near the superconductor-Mott
insulating phase transitions.
\cite{Glazman}.

  In this letter, we investigate a new design of the Josephson
junction by taking advantage of the  competition between the
superconducting and Mott-insulating phases.
The two sides of the junction are hole and electron-doped
superconductors, respectively, which are close to the
superconductor-Mott-insulator transition. At the interface, a
self-organized Mott-insulating region is formed as the tunneling
barrier.  We dub this junction ``{\it Josephson diode}''. Similarly
to the conventional $p$-$n$ junctions in semiconductor diodes, the
depletion region is suppressed by a positive bias voltage and
elongated by a negative bias voltage, giving rise to an asymmetric
response of the Josephson current to the external voltage.
While the geometry of the junction is
similar to the  $p$-$n$ junctions in semiconductor diodes,
there are fundamental differences.
In the Josephson diode, the depletion region (Mott-insulator)
is formed completely due to the quantum nature of competing phases.
Moreover, the depletion region
not only plays a role of tunneling barrier, but also its quantum
fluctuation reveals important information of strong correlation
effects. The fluctuation of the region boundaries couples to the
carrier recombination process which results in an additional phase
dynamics that can be measured experimentally in the noise spectrum
of the  Josephson current.

\begin{figure}
\includegraphics[width=6cm,height=3cm]{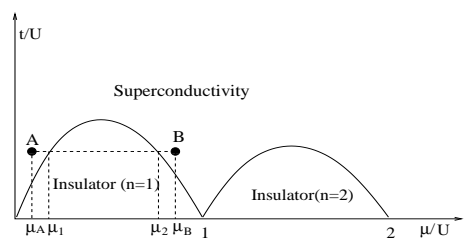}
\caption{\label{fig1:phase} The mean-field phase diagram of the
Bose-Hubbard model. The left (right) side of the junction
corresponds to the states at point  $A (B)$. }
\end{figure}

To illustrate the physics of the Josephson diode, we use
the Bose-Hubbard model for charged bosons.
The long range interaction between bosons is approximated by
the self-consistent effective potential $V$  as
\begin{eqnarray}
H&=&-t\sum_{\langle ij \rangle}
(b^\dagger(i) b(j)+h.c.)+\frac{U}{2}\sum_{i}n (i)(n(i)-1)\nonumber\\
&+&\sum_{i} (V(i)-\mu) b^\dagger(i) b(i),
\label{eq:BHubbard}
\end{eqnarray}
 where $t$ is the hopping integral, $U$ is the short  range on-site
repulsion. In homogeneous systems,  $V(i)$ equals zero due to the
charge neutrality maintained by the background charge,
and the phase diagram is well-known as  shown in Fig.
\ref{fig1:phase}. Mott-insulating phases appear at commensurate
fillings and small values of $t/U$, which can be doped into
superconducting phases either by particles or holes.

The structure of the Josephson diode is depicted in Fig. \ref{fig2} a:
its left and right hand sides correspond to two points $A$ and $B$
in Fig. \ref{fig1:phase}, where the particle densities $\langle n(i) \rangle_0
=1 \mp \delta$ for $i$ on the left (right) side, respectively.
Due to the chemical potential imbalance, bosons diffuse from the right to
left side.
Since a Mott-insulating region interpolates between the hole-doped
superconductor $A$ and the particle-doped superconductor $B$,
it appears across the junction in the real space as the depletion
region depicted in Fig. \ref{fig2} b.
We define $\mu_{1(2)}$ as the chemical potentials at the boundary
of the Mott-insulating phase depicted in Fig. \ref{fig1:phase}.
Inside this region, the density $\langle n(i)\rangle$ is fixed to
the commensurate value $1$
and  the charge neutrality is no longer kept.
The resulting internal electric field $E$ gives rise a change
of the electric potential $\Delta V$ across the junction to compensate the
local chemical potential difference.
As a result, the total chemical potential $\mu$ becomes constant.
The $E$ field is given by
\begin{eqnarray}
\nabla E(i)= -\nabla^2 V(i) = \frac{q}{\epsilon}(n(i)-\langle n(i) \rangle_0),
\label{eq1}
\end{eqnarray}
where $q$ is the charge of bosons, $\epsilon$ is the dielectric
constant and the lattice constant is taken as 1.
In the following, we consider the case that
$\Delta \mu=\mu_2-\mu_1$ is much larger than $\mu_1-\mu_A$ and $\mu_B-\mu_2$.
In this case, most of the charge non-neutral region is the Mott-insulating,
and we neglect the contribution to the $E$ field outside the Mott-insulating
region.
Then $\Delta V$ across the depletion region can be approximated as
$\Delta V \approx -\int^{D}_{-D} E(x) d x=
\frac{D^2 q \delta}{\epsilon}=\Delta \mu$,
where $2D$ is the length of the Mott-insulating depletion region,
thus
\bea
D\approx  \sqrt{ \frac{\epsilon \Delta \mu}{q \delta }  }.
\label{eq:length}
\eea

\begin{figure}
\includegraphics[width=6cm, height=5cm]{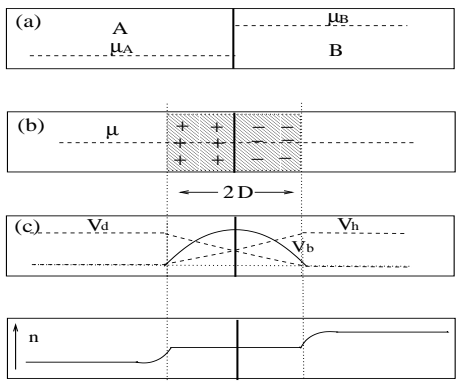}
\caption{\label{fig2} The sketch of the junction: (a) the spatial
distribution of chemical potential before equilibrium is
formed; (b) the Mott-insulating depletion region and  charge distribution
in equilibrium; (c) the potentials $V_h$ and $V_d$ for the holons
and doublons  across the junction, respectively. 
}
\end{figure}

Next we discuss the Josephson diode effect under an external voltage $V_{ex}$.
In the conventional Josephson junction, the coupling term  is given by
\begin{eqnarray}
H_J= -J \cos(\phi_r-\phi_l+\frac{q V_{ex} t}{\hbar}),
\end{eqnarray}
where the coupling strength $J$ is not sensitive to $V_{ex}$.
However, the situation in the Josephson diode junction is quite different.}
When $V_{ex}$ is applied, it not only creates the time dependent
phase difference across the junction, but also changes
the length of the depletion region.
Similarly to the case of semiconductor diodes,
the depletion region is suppressed (elongated)
in the forward (reversed)-biased junction since the external electric field
aligns in the opposite (same) direction of the internal electric field.
By a similar reasoning for Eq. \ref{eq:length}, we obtain
$D$ in the presence of $V_{ex}$ as
$D\approx \sqrt{ \epsilon (\Delta \mu-V_{ex})/(q \delta)   }$.
When $D$ is much longer than the coherent length $\xi$, the Josephson
coupling $J$ is proportional to $e^{-\alpha\frac{D}{\xi}}$,
where $\alpha$ is a dimensionless constant.
Therefore, $J$ as a function of $V_{ex}$ is asymmetric as
\bea
\log J(V_{ex}) \propto \sqrt{(\Delta \mu-V_{ex})/(q \delta)}.
\label{eq:jcoupling}
\eea
which is sketched in Fig. \ref{fig3}.
This is a natural generalization of the
effect of the $p$-$n$ junction in semiconductors.
Please note that when $V_{ex}$ is close to $\Delta \mu$, the Mott depletion
region is almost completely suppressed, then Eq. \ref{eq:jcoupling}
does not apply.

A major difference between the Josephson and  conventional
semiconductor diodes is that the charge current is no longer
unidirectional in the former case. In the semiconductor diode, the
unidirectional charge transport clearly breaks time reversal
symmetry, which is consistent with its dissipative nature. In
contrast, in the Josephson diode, the dissipationless transport
keeps time reversal symmetry. As a result, in the zero bias case,
the DC Josephson current can flow either from the $p$ to  $n$-side
or equivalently well from the $n$ to $p$-side. At the finite bias,
the AC Josephson current flows back and forth between the two sides.
The asymmetric response to the external voltage lies in the
magnitude of the Josephson critical current, which is an effect from
the explicit parity breaking in the structure of the Josephson diode
junction.


\begin{figure}
\psfrag{V}{$V_{ex}$}
\psfrag{J(V)} {$J(V_{ex})$}
\psfrag{D(V)} {$D(V_{ex})$}
\includegraphics[width=5cm, height=2.5cm]{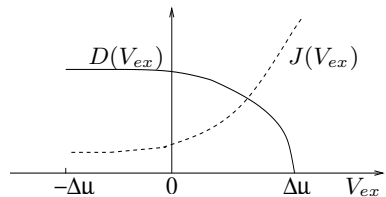}
\caption{ The length of Mott-insulating depletion region $D$ and the
Josephson coupling strength $J$ as a function of the external
voltage $V_{ex}$ ( arbitrary unit for the vertical axis). }
\label{fig3}
\end{figure}

In a semiconductor $p$-$n$ junction, there are the processes of the
diffusion and recombination of  charge carriers.
The analog of such processes also exists
in the Josephson diode junction.
We introduce the singlon ($s^\dagger (i)$), holon ($a_{h}^\dagger (i)$),
and doublon ($a_{p}^\dagger (i)$) operators for the empty, single and
double-occupied states  on the site $i$, respectively.
We limit the Hilbert space on each site to these three states,
which is a good assumption for the Bose-Hubbard model when
the density is close to one.
In this case, the holons and doublons are considered as
the majority carriers in the left and right sides, respectively.
Since each site can only have one such boson, these bosons are
hard core-like, thus namely they satisfy the following constraint $
a^\dagger_h a_h+s^\dagger s + a_p^\dagger a_p =1. $
We can rewrite the original boson operators in terms of these hard core bosons
as 
$ b^\dagger= s^\dagger a_h+a_p^\dagger  s. $
The density operator of the original bosons is given by
$ n=b^\dagger b=a_p^\dagger a_p-a_h^\dagger a_h+1. $
The hopping term in this representation becomes
\begin{eqnarray}
H_t &=& -t\sum_{ij} \Big\{ [
a^\dagger_{h}(j)a_{h}(i)
+a^\dagger_{p}(i)a_{p}(j) ] s^\dagger(j)s(i) +h.c.
 \nonumber
\\& +&[a_{h}(i)a_{p}(j)+
a_{p}(i)a_{h}(j)]
s^\dagger(i)s^\dagger(j)+h.c. \Big\},
\label{eq:hop}
\end{eqnarray}
where the first two terms describe the hopping processes of holons and doublons
and the second two terms describe their recombination and  regeneration
processes.
Correspondingly, the current operator $J_{cur}$ is given by
\begin{eqnarray}
& &J_{cur}  =  it\sum_{ij}\{
[a^\dagger_{h}(i)a_{h}(j) - a^\dagger_{p}(j)a_{p}(i)]
s^\dagger(i)s(j)-h.c.\}\nonumber \\
& & -it\sum_{ij}\{ [a_{h}(i)a_{p}(j)-a_{h}(j)a_{p}(i))s^\dagger (j)s^\dagger
(i)-h.c.\},
\end{eqnarray}
where the first two terms are the drift currents and the last two terms are
recombination and regeneration current.



In order to study the Josephson coupling explicitly in the diode,
we take the following  general mean field trial wavefunction,
 \begin{eqnarray}
 \Psi & = &\prod_i \Big\{ \cos\theta_1(i) s^\dagger(i) +  e^{i\phi_2(i)}
\sin\theta_1(i) [ e^{i\phi_1(i)}
\nonumber \\
& & \cos\theta_2(i)a_h^\dagger(i) + e^{-i\phi_1(i)} \sin\theta_2(i)
a_p^\dagger(i) ] \Big\},
\end{eqnarray}
which gives the superconducting order parameter as
 \bea
\avg{b^\dagger (i)}&=& \frac{\sin 2\theta_1(i)}{2}e^{i\phi_1(i)}
(e^{i\phi_2(i)}\cos\theta_2(i)\nonumber \\
&+&e^{-i\phi_2(i)}\sin\theta_2(i)).
 \eea
There are two independent phase variables $\phi_{1,2}$ in the above formulae.
$\phi_1(i)$ is the conventional phase of
superconducting order parameter, which is the conjugate variable
to the density, namely, $ [n(i), \phi_1(i)]=i$.
The other variable,  $\phi_2(i)$ is associated with the quantity
$a_p^\dagger a_p+a_h^\dagger a_h = 1-s^\dagger s$, namely, the
density of singlons  $n_s(i)= s^\dagger s$.
Thus, we have $ [n_s(i), \phi_2(i)]=i.$
In the conventional Josephson junction,
$\phi_2(i)$ can be simply fixed to be zero since the number
of singlons can  arbitrarily fluctuate.
However, the presence of the depletion insulating
region changes the above picture.
Not only does the depletion region serve as a tunneling barrier,
but also its fluctuation leads to additional physics, which
is the fluctuation of the number of singlons.
Assuming the variables of the wavefunctions in the superconducting
states in the left and right sides of the junctions are given by
$(\phi_{1l,r},\phi_{2l,r},\theta_{1l,r},\theta_{2l,r})$, we
rewrite Eq. \ref{eq:hop} as
\begin{eqnarray}
H_t&=& J_1 \cos(\Delta\phi_1- \Delta\phi_2)+J_2 \cos(\Delta\phi_1+
\Delta\phi_2)
\nonumber \\
&+&J_3 \cos(\Delta\phi_1- \varphi)+J_4 \cos(\Delta\phi_1+ \varphi),
\label{eq:hidden} \\
\Delta\phi_i&=&\Delta\phi_{i r}-\Delta\phi_{i l} ~(i=1,2), \ \ \
\varphi=\phi_{2l}+\phi_{2r},
\nonumber\\
J_1&=&\sin 2\theta_{1l} \sin 2\theta_{1r} \cos \theta_{2l}
\cos \theta_{2r}/4, \nonumber \\
J_2&=&\sin 2\theta_{1l} \sin 2\theta_{1r} \sin \theta_{2l} \sin \theta_{2r}/4,
\nonumber \\
J_3&=& \sin 2\theta_{1l} \sin 2\theta_{1r} \cos \theta_{2l} \sin \theta_{2r}/4,
\nonumber \\
J_4&=& \sin 2\theta_{1l} \sin 2\theta_{1r} \sin \theta_{2l} \cos \theta_{2r}/4.
\eea

Similarly to the Josephson coupling in multi-band superconductors
\cite{leggett,agterberg,leggett2}, Eq. \ref{eq:hidden} indicates
hidden internal dynamics in the junction. There are two
observations. The first is that the coupling depends not only on
$\Delta\phi_{2}$, but also on $\varphi$, the sum of the phases
$\phi_{2l}$ and $\phi_{2r}$, reflecting the recombination and
regeneration processes.  As shown in Fig. \ref{fig1:phase}, holons
and doublons are the majority carriers in the left and right sides
respectively, thus $\cos\theta_{2l}\gg \sin\theta_{2l}$ and $\sin
\theta_{2r}\gg \cos\theta_{2r}$ hold. As a result, the $J_3$ term
describing the recombination and regeneration processes dominates
over other terms. The second is that if $\phi_2$ is set to $0$, we
recover the conventional form of the Josephson coupling energy i.e.
$H=J \cos(\Delta\phi_1)$.
Without the insulating region, the number of 
singlons is not determined in both sides. Therefore, $\varphi$ and $\Delta
\phi_2$ do not have any dynamics and can be chosen to 0.
However,  their dynamics can be created in
the presence of the depletion insulating region.

We discuss the fluctuation of the Mott-insulating depletion region.
When the system is slightly out of equilibrium, the depletion region
has to shrink or expand symmetrically respect to the center of the
junction in order to maintain the charge neutrality.
This corresponds to the recombination and regeneration processes
in the depletion region, which changes the value of $n_s^R+n_s^L$,
and thus creates the dynamics of $\varphi$.
The frequency of this mode can be estimated as follows.
The mismatch of chemical potentials in the two sides of the junctions
depends on the length of the depletion region as
$\Delta\mu_{mis}=\Delta \mu-\Delta V=\Delta\mu- D^2 q \delta/\epsilon$.
The restoration force on the charge carriers can be approximated as
$F=-\Delta\mu_{mis}/(2D)=-\Delta x \Delta \mu/D^2$, where $\Delta x$ is the small displacement of the boundaries of the depletion region.
This mode is essential the plasmon mode with the frequency determined as
$\omega_b^2=q\delta/(m^* \epsilon)$, where $m^*\approx 1/(t\delta)$ is
the effective mass of the charge carrier.
Thus $\omega_b$ scales linearly with the doping level $\delta$.

In general, the fluctuations of the phase $\varphi$ is a diffusion
process. Let $\Gamma$ be the decay rate set by the diffusion, which
depends on disorder and detailed low energy excitations in the
insulating and superconducting regions.
The fluctuation of $\varphi$ lead to the following correlation function,
\begin{eqnarray}
\avg{e^{i\varphi(t)}e^{-i\varphi(0)}} & = &e^{i\omega_b t-\Gamma |t|}
\end{eqnarray}
Therefore, the total current correlation function ,
\begin{eqnarray}
\avg{J(t)J(0)}
 &\propto &\cos(\omega_b t)e^{-\Gamma |t|},
\end{eqnarray}
which gives  the noise spectrum of current as
\begin{eqnarray}
S(\omega) &=& \int dt e^{i\omega t} \avg{\{J(t)-\avg{J(t)}, J(0)-\avg{J(0)}\}}
\nonumber \\ &\propto&
\frac{\Gamma}{(\omega-\omega_b)^2+\Gamma^2}
+\frac{\Gamma}{(\omega+\omega_b)^2+\Gamma_2^2} .
\end{eqnarray}

Now we briefly discuss the case for neutral bosonic system, say, the
$p$-$n$ junction made by cold bosonic atoms. The Bose-Hubbard model
\cite{Greiner} has already been realized in optical lattices
experimentally. In such systems, it requires an external potential
drop to create a depletion region in an inhomogeneous  optical
lattice. In this neutral system, there is an additional mode, a
sliding mode, associated to the collective excitations of depletion
region. For the sliding mode, we can imagine that the system is
slightly out of equilibrium by sliding the whole depletion region.
Since this mode changes the value $(n_s^R-n_s^L)$, it naturally
creates the dynamics of $\Delta \phi_2$. In the two band picture,
this mode is corresponding to transfering a hole (or doubleon) in
the one side to the other side. Since the difference of band energy
is given by the mott gap, $\Delta \mu$ which is proportional to $U$,
the mode is expected to have the frequency $\omega_a=\Delta \mu$.
This mode has the same response in the  noise spectrum of the
junction as the plasma mode expect their energy difference.  A
design of the ``atomtronic diode'' exhibiting asymmetric response to
external chemical potential difference was discussed by Seaman {\it
et al.} \cite{Seaman}. The major difference to our design is that no
insulating depletion region  exists in their design.


Experimentally, Bose-Hubbard mode for charged bosons can be realized
in Josephson arrays \cite{Glazman}.
By choosing the chemical potential configuration discussed above,
the Josephson diode should be formed.
High T$_c$ cuprates are another class of possible systems to realized the
Josephson diode.
Two different insulating phases exist in cuprates.
One is in undoped systems and the other is at the 1/8
doping level \cite{kivelson,zaanen}, both of which
can be used to create the Josephson diode in principle.
A universal asymmetric density between positive and negative bias
voltage in STM experiments \cite{hoffman,davis,arli} has been observed,
which can been explained by considering an insulating competing order
at surface \cite{hu,sudip,DiStasio,chakravarty1}.
This is an important support for the possible realization of the
proposed junction, although other alternative explanations
\cite{anderson,wen} also exist.
The key to create the Josephson diode junction is a spatially selective
doping control in cuprates.
An in-plane SAS junction has been created by using the spatially selective
and reversible doping control techniques in cuprate films \cite{Oh}.
Photodoping techniques can also locally control the doping level
\cite{Gilabert,Tanabe, Decca}.
These techniques can be used to create two close-by regions with
different doping levels with an insulating phase between them.
Although it is clear that the Josephson diode in cuprates can not
be described by the simple Bose-Hubbard model, we believe that
the physics presented here is still valid.
The effective bosonic Hamiltonian in cuprates has to include the spin
degree of freedom \cite{Eugene, project,Tsai}.
Since we are interested in the charge transport in the Josephson diode,
the spin degree of freedom does not have direct effect on the
properties we studied in this paper.
However, it is largely an open question whether any other new
properties can be induced by the spin degree of freedom.

We would like to acknowledge the earlier stimulating conversations
with S. C. Zhang. J. P. Hu would like to thank the summer school at
Boulder, Colorado. Most of the work in the paper was reported in
poster session in this summer school in 2003. J.P. Hu was supported
by National Scientific Foundation under award number Phy-0603759.

\end{document}